\documentclass[preprint,pre,amsmath,amssymb,showpacs]{revtex4}

\usepackage{epsfig}
\usepackage{epstopdf}
\usepackage{graphics}
\usepackage{multirow}
\usepackage{bm}
\begin{document}
\title{Kinetic theory based force treatment in lattice Boltzmann equation}
\author{Lin Zheng$^{1}$}
\email[Corresponding author:\quad]{lz@njust.edu.cn}
\author{Song Zheng$^2$}
\author{Qinglan Zhai$^3$}
\affiliation{1 MIIT Key Laboratory of Thermal Control of Electronic Equipment, School of Energy and Power Engineering, Nanjing University of Science and Technology, Nanjing 210094, P.R. China}
\affiliation{2 School of Mathematics and Statistics, Zhejiang University of Finance and Economics, Hangzhou 310018, P.R. China}
\affiliation{3 School of Economics Management and Law, Chaohu University, Chaohu 238000, P.R. China}
\pacs{47.11.-j}
\begin{abstract}
In the gas kinetic theory, it showed that the zeroth order of the density distribution function $f^{(0)}$ and local equilibrium density distribution function were the Maxwellian distribution $f^{(eq)}(\rho,\bm u, T)$ with an external force term, where $\rho$ the fluid density, $\bm u$ the physical velocity and $T$ the temperature, while in the lattice Boltzmann equation (LBE) method numerous force treatments were proposed with a discrete density distribution function $f_i$ apparently relaxed to a given state $f^{(eq)}_i(\rho,\bm u^*)$,  where the given velocity $\bm u^*$ could be different with $\bm u$, and the Chapman-Enskog analysis showed that $f^{(0)}_i$ and local equilibrium density distribution function should be $f^{(eq)}_i(\rho,\bm u^*)$ in the literature. In this paper, we start from the kinetic theory and show that the $f^{(0)}_i$ and local equilibrium density distribution function in LBE should obey the Maxwellian distribution $f^{(eq)}_i(\rho,\bm u)$ with $f_i$ relaxed to $f^{(eq)}_i(\rho,\bm u^*)$, which are consistent with kinetic theory, then the general requirements for the force term are derived, by which the correct hydrodynamic equations could be recovered at Navier-Stokes level, and numerical results confirm our theoretical analysis.
\end{abstract}
\maketitle

\section{Introduction}

The fluid transport phenomena is the results of the molecular random motions or interactions at microscopic level and it is observed in nature or engineering processes. From the kinetic theory, the molecular random motions could be described by a probability density distribution function $f(\bm x, \bm\xi, t)$ at position $\bm x$ and time $t$ with molecular velocity $\bm\xi$, and its evolution equation is governed by Boltzmann equation (BE), which could capture the transport phenomena in all fluid regimes \cite{Cercignani}. Owing to the complex of the integro-differential collision operator, it's difficult to derive a solution directly. Fortunately, some attractive discrete methods are developed to mimic the BE with a Bhatnagar-Gross-Krook (BGK) collision operator \cite{BGK} such as lattice Boltzmann equation (LBE) method \cite{Succi,GuoB} and gas kinetic scheme \cite{Xu}, which show a great successful application in computational fluid dynamics \cite{Xu,Aidun,Mehr,Torres}.

As we know, Chapman-Enskog (CE) analysis is widely used in the gas kinetic theory, and it is also an important tool in LBE. There has no doubt about the CE analysis for the discrete density distribution function $f_i$ without a force term in LBE, in which $f^{(0)}_i$ is determined by local Maxwellian equilibrium density distribution $f^{(eq)}_i(\rho,\bm u)$ with fluid density $\rho$ and physical velocity $\bm u$. However, when a force is involved in a fluid system, the corresponding force term is usually used to incorporate its effect in the LBE, and the formulations of $f^{(0)}_i$ and local Maxwellian equilibrium density distribution are not reached the unified viewpoint. In the literature, some classical force treatments such as Shan-Chen scheme \cite{Shan,Shan1}, He \emph{et al} scheme \cite{He} Ladd's scheme\cite{Ladd}, Guo \emph{et al.} scheme \cite{Guo1}, Wagner's scheme \cite{Wagner} and the exact-difference-method (EDM) scheme \cite{Kuper} were used to include the force in LBE community. The aforementioned force schemes could be written in the unified formulation denoted by $F_i$ and $f_i$ was apparently relaxed to $f^{(eq)}_i(\rho,\bm u^*)$ with a relaxation time $\tau$ and $\bm u^*$ as a given velocity, that is, a BGK-like relaxation process was applied to model the collision operator term. In Refs. \cite{Ladd,Guo1}, the results of CE analysis showed $f^{(0)}_i$ was equal to $f^{(eq)}_i(\rho,\bm u^*)$, where they took $f^{(eq)}(\rho,\bm u^*)$ as the local Maxwellian equilibrium density distribution function during the analysis, while the kinetic BE required both $f^{(0)}$ and the local equilibrium density distribution function should be the Maxwellian distribution $f^{(eq)}(\rho,\bm u, T)$. In this paper, we aim to address these issues from the kinetic theory and analyze the consistency between LBE and the kinetic BE with a force term.

To this end, we start from the kinetic theory, and analyze the local equilibrium density distribution function and the zeroth order distribution function $f^{(0)}$ by CE, then derive a relation between the LBE with an apparently relaxation process and BE. The rest of this paper is organized as follows. In Sec. II, the consistency between LBE and BE is analyzed in detail, to recover the correct Navier-Stokes (NS) equations, the general requirement of force term in LBE is presented, then numerical simulations are conducted to validate our analysis in Sec. III, and finally a brief conclusion is given in Sec. IV.
\section{Kinetic theory consistency with a force term }
From the kinetic theory, the simplified BE with the BGK collision operator could be written as \cite{BGK}
\begin{equation}
\partial_t f+\xi_\alpha\partial_\alpha f+a_\alpha\cdot\nabla_{\xi_\alpha}f=-\frac{f-f^{(eq)}(\rho,\bm u, T)}{\tau}=\Omega_{BGK}(f),
\label{Eq1}
\end{equation}
where $\xi_\alpha$ is the molecular velocity, $a_\alpha$ is an acceleration. Obviously, $\Omega_{BGK}$ satisfies the following conservation conditions:
\begin{equation}
\int\psi_\alpha\Omega_{BGK}(f)d\xi_\alpha=0,
\label{Eq2}
\end{equation}
with $\psi_\alpha$=1, $\xi_\alpha$, $\xi^2/2$ as collision invariants, and the local equilibrium density distribution function $f^{(eq)}(\rho,\bm u,T)$ in Eq. (\ref{Eq1}) is to be
 \begin{equation}
f^{(eq)}(\rho,\bm u, T)=\frac{\rho}{(2\pi RT)^{D/2}}\texttt{exp}\left[\frac{|\bm\xi-\bm u|^2}{2RT}\right],
\label{Eq3}
\end{equation}
where $R=k_b/m$ is gas constant with $k_b$ the Boltzmann constant and $m$ the molecular mass, $D$ is the dimension, $T$ is the temperature. The fluid density $\rho$, physical velocity $\bm u$ and temperature $T$ are defined by
 \begin{equation}
\rho=\int fd\bm\xi,~~~~\rho\bm u=\int\bm\xi fd\bm\xi,~~~~\frac{\rho DRT}{2}=\int\frac{|\bm\xi-\bm u|^2}{2} fd\bm\xi.
\label{aEq3}
\end{equation}

Now we apply the CE analysis to the BE, that is, $\partial_t=\sum\limits_{k=0}^\infty\epsilon^k\partial_{t_k}$, $f=\sum\limits_{k=0}^\infty\epsilon^k f^{(k)}$, and $\Omega_{BGK}(f)=\epsilon^{-1}\Omega_{BGK}(f)$, then Eq. (\ref{Eq1}) can be written in consecutive orders of $\epsilon$ as
\begin{equation}
\sum^{\infty}_{n=0}\sum^{n}_{k=0}\epsilon^{(n+1)}D_nf^{(n-k)}=\Omega_{BGK}(f^{(0)})+\sum^{\infty}_{n=0}\sum^{n}_{k=0}\epsilon^{(n)}\Omega_{BGK}(f^{(k)})
\label{Eq4}
\end{equation}
where $D_n=\partial_{t_n}+\xi_\alpha\nabla_\alpha+a_\alpha\nabla_{\xi_\alpha}$, and the CE analysis shows that $f^{(0)}$ is equal to the local equilibrium density distribution function $f^{(eq)}(\rho,\bm u,T)$.

If $f$ is relaxed to a given state $f^{(eq)}(\rho, \bm u^*, T)$ with $\Omega_{gs}=-(f-f^{(eq)}(\rho,\bm u^*, T))/\tau$ in order to model the collision term in BE, the following evolution equation could be derived
\begin{equation}
\partial_t f+\xi_\alpha\partial_\alpha f+a_\alpha\cdot\nabla_{\xi_\alpha}f=\Omega_{gs},
\label{Eq6}
\end{equation}
however, the basic properties of the collision invariants conditions for $\Omega_{gs}$ can not be guaranteed with $\psi_\alpha\neq 1$ and $\bm u^*\neq\bm u$
 \begin{equation}
\int\psi_\alpha\Omega_{gs}d\xi_\alpha\neq\int\psi_\alpha\Omega_{BGK}d\xi_\alpha=0,
\label{Eq7}
\end{equation}
this implies that the well-known H-theorem $dH/dt=\int(1+\texttt{ln}f)\partial_t fd\xi_\alpha dx_\alpha\leq0$ can not be guaranteed by Eq. (\ref{Eq6}), and the local equilibrium density distribution function can not be $f^{(eq)}(\rho,\bm u^*, T)$ with $\bm u^*\neq\bm u$. Therefore, the given state in the BGK-like relaxation process must be $f^{(eq)}(\rho,\bm u, T)$ to model the collision operator term in BE, otherwise, the solution of Eq. (\ref{Eq6}) will deviate from the original BE. In the literature, many LBE models were developed to incorporate the force effect \cite{Shan,Shan1,He,Ladd,Guo1,Wagner,Kuper}, and the appearance of collision operator was similar to $\Omega_{gs}$ in BGK-like relaxation process, but the underlying physics was not clarified. To this end, we begin with the general evolution equation which can be written as
\begin{equation}
\partial_t f+\xi_\alpha\partial_\alpha f+F=\Omega_{gs},
\label{Eq8}
\end{equation}
where $F$ is the force term to be determined latter. As afore analysis, if the given state relaxation process is used to model the collision operator in BE, the given state must be $f^{(eq)}(\rho,\bm u,T)$. Therefore, Eq. (\ref{Eq8}) should be rewritten as
\begin{equation}
\partial_t f+\xi_\alpha\partial_\alpha f+G=\Omega_{BGK},
\label{Eq9}
\end{equation}
where $G=F+(f^{(eq)}(\rho,\bm u,T)-f^{(eq)}(\rho,\bm u^*,T))/\tau$. To be consistent with the simplified BE in Eq. (\ref{Eq1}), the general formulation of $F$ should be chosen
\begin{equation}
F=a_\alpha\nabla_{\xi_\alpha}f -\frac{f^{(eq)}(\rho,\bm u,T)-f^{(eq)}(\rho,\bm u^*,T)}{\tau},
\label{Eq10}
\end{equation}
with Eq. (\ref{Eq10}) and the CE analysis to Eq. (\ref{Eq8}), it is shown that both $f^{(0)}$ and the local equilibrium density distribution function in Eq. (\ref{Eq8}) are $f^{(eq)}(\rho,\bm u,T)$.

Integrating (\ref{Eq8}) along the characteristic lines as \cite{He}, it gives the following discrete evolution equation
\begin{equation}
\bar f_i(\bm x+\bm\xi_i\delta t)- \bar f_i(\bm x, t)=-\omega_f(\bar f_i(x, t)-f^{(eq)}_i(\rho, \bm u^*))+\delta t(1-\frac{\omega_f}{2})F_i,
\label{Eq11}
\end{equation}
where $\bar f_i=f_i-\delta t(\Omega_{gs}+F_i)/2$
with $i$ as the $i$th direction of the discrete velocity, time increment $\delta t$, and $\omega_f=2\delta t/(2\tau+\delta t)$, and the discrete formulation of the given state $f^{(eq)}_i(\rho, \bm u^*)$ is
\begin{equation}
f^{(eq)}_i(\rho,\bm u^*)=\omega_i\rho\left[1+\frac{\xi_{i\alpha}u^*_\alpha}{RT}+\frac{(\xi_{i\alpha}\xi_{i\beta}-RT\delta_{\alpha\beta})u^*_\alpha u^*_\beta}{2RT^2}\right],
\label{Eqa12}
\end{equation}
with $\omega_i$ as the corresponding weight coefficient to the discrete velocity set, then the fluid density $\rho$ and physical velocity $\bm u$ are defined by
 \begin{equation}
\rho=\sum_i\bar f,~~~~\rho\bm u=\sum_i\bm\xi_i \bar f_i+\frac{\delta t}{2}\bm F.
\label{Eq13}
\end{equation}

To derive the correct hydrodynamic equations at NS level, as long as $F_i$
satisfies the following conditions
\begin{eqnarray}
\label{Eq14}
\sum_iF_i=0,~~B_\alpha=\sum_i\xi_{i\alpha}F_i=F_\alpha-\frac{\rho(u_\alpha-u^*_\alpha)}{\tau},\\\nonumber
C_{\alpha\beta}=\sum_i\xi_{i\alpha}\xi_{i\beta}F_i=u_\alpha F_\beta+F_\alpha u_\beta-\frac{\rho(u_\alpha u_\beta-u^*_\alpha u^*_\beta)}{\tau},
\end{eqnarray}
and the corresponding discrete formulation of $F$
at Navier-Stokes level can be written as
\begin{equation}
F_i=\omega_i\left[\frac{\xi_{i\alpha}B_\alpha}{RT}+\frac{(\xi_{i\alpha}\xi_{i\beta}-RT\delta_{\alpha\beta})C_{\alpha\beta}}{2RT^2}\right].
\label{Eq16}
\end{equation}

From the afore theoretical analysis, it should be stressed that the collision term in Eq. (\ref{Eq11}) seems to be $-\omega_f(\bar f_i-f^{(eq)}_i(\rho, \bm u^*))$, but physically it can not be termed as "collision term" which is modeling the original collision operator in BE, in fact, it includes the physical collision term $-\omega_f(\bar f_i-f^{(eq)}_i(\rho, \bm u))$, which is derived from Eq. (\ref{Eq9}) to model the collision operator in BE. This implies that $-\omega_f(\bar f_i-f^{(eq)}_i(\rho, \bm u^*))$ could be decomposed into two parts: one as the physical collision term and another as a deviation term, and the CE analysis shows that both $f^{(0)}_i$ and the local equilibrium density distribution function in Eq. (\ref{Eq11}) are $f^{(eq)}_i(\rho, \bm u)$, which are consistent with kinetic theory.

If the derivation of $\bm u^*$ to $\bm u$ is assumed to be $\delta\bm u$, that is, $\bm u^*=\bm u+\delta \bm u$, we can prove from Eqs. (\ref{Eq8})-(\ref{Eq11}) that as long as $F_i$ satisfying Eq. (\ref{Eq14}) any value of $\delta\bm u$ could give the same predictions by Eq. (\ref{Eq11}).  Obviously, $\delta \bm u=0$ gives the Guo \emph{et al.} force scheme \cite{Guo1} and $\delta \bm u=-\delta t\bm F/2\rho$ gives Wagner's force scheme \cite{Wagner}, in which the high order interfacial term is neglected. Both schemes meet the required conditions in Eq. (\ref{Eq14}), and theoretically they should give the same predictions.

\section{Numerical simulations}

In this section, three benchmark problems are carried out to validate our theoretical analysis by two-dimensional nine discrete velocity (D2Q9) LBE, \emph{i.e.}, $\bm\xi_0=(0,0)$, $\bm\xi_{i=1-4}=(\cos[(i-1)\pi/2],\sin[(i-1)\pi/2])$, $\bm\xi_{i=5-8}=\sqrt{2}(\cos[(2i-9)\pi/4],\sin[(2i-9)\pi/4])$, and the corresponding weight coefficients in Eqs. (\ref{Eqa12}) and (\ref{Eq16}) are $\omega_0=4/9$, $\omega_{1-4}=1/9$ and $\omega_{5-8}=1/36$. The first test problem is a two-dimensional Taylor-vortex flow driven by a time and space dependent external force, then a stationary droplet immersed to another fluid  is investigated by pseudopotential LBE and finally is a droplet on wettability solid.

\subsection{Taylor vortex flow}
\begin{figure}
\includegraphics[width=2.5in,height=2.0in]{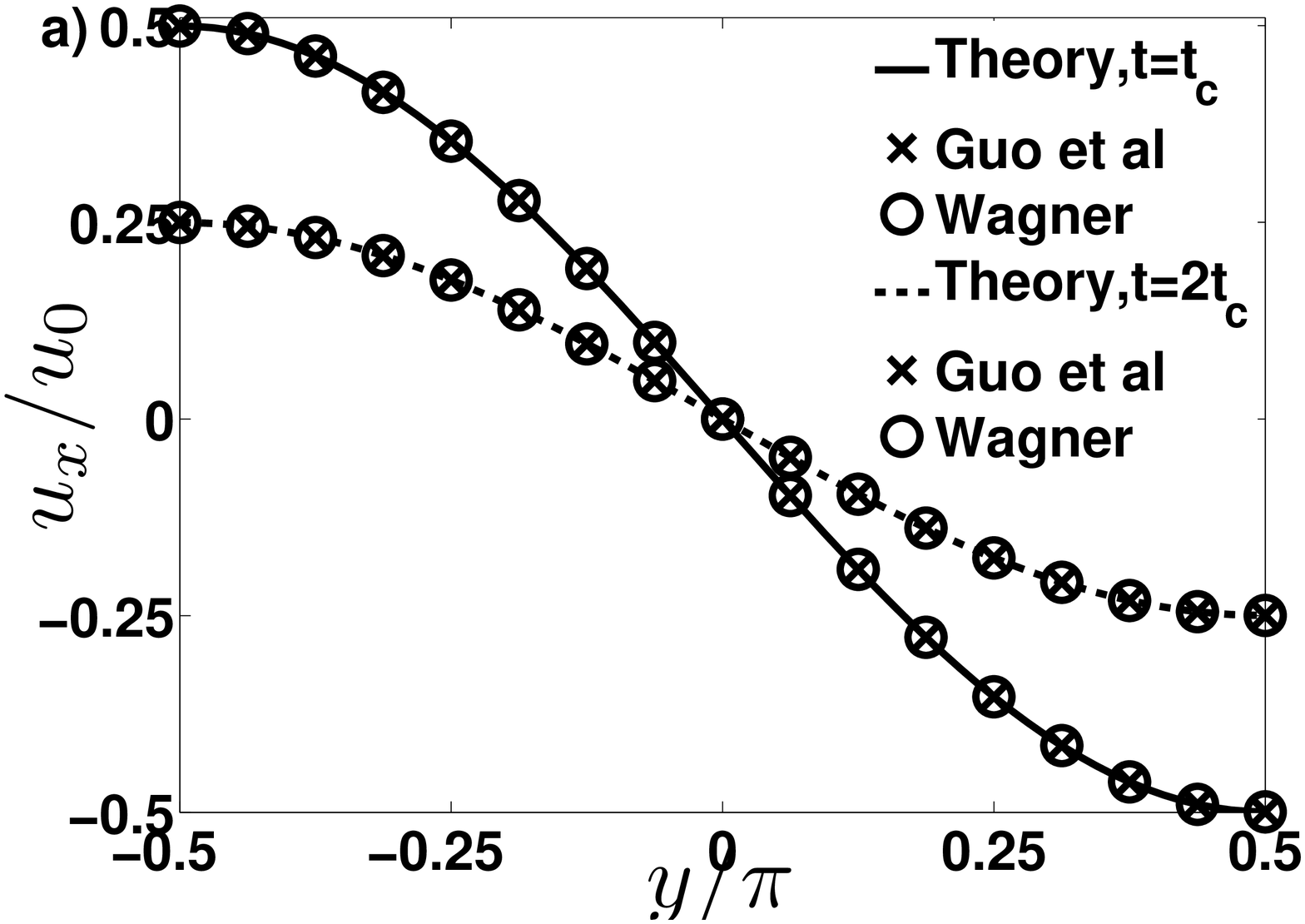}%
\includegraphics[width=2.5in,height=2.0in]{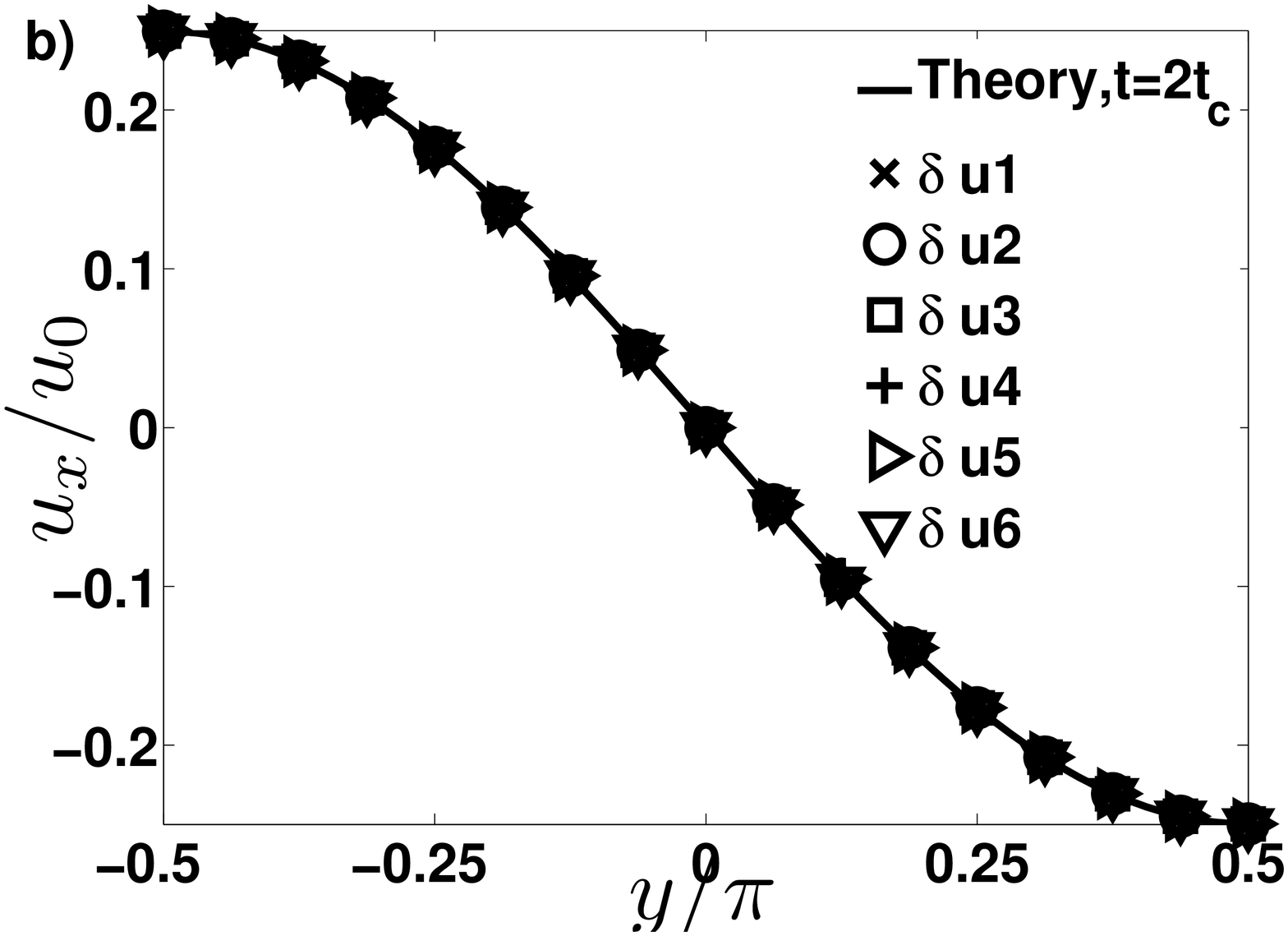}
\caption{Numerical predictions of $u_x(0,y*)$. a) the comparison of Guo et al scheme \cite{Guo1} and Wagner's scheme \cite{Wagner} at $t=t_c$ and $2t_c$; b) the predictions by different $\delta \bm u$ at $t=2t_c$.}\label{Fig1}
\end{figure}

In Fig. 1
, a two-dimensional unsteady Taylor vortex problem is investigated by LBE with force schemes of Guo \emph{et al.} and Wagner, in this case, the flow is driven by an external force $\bm F=(F_x, F_y)=(-k_1\rho u^2_0\texttt{sin}(2k_1x)\texttt{exp}[-2\nu(k^2_1+k^2_2)t]/2,-k^2_1\rho u^2_0\texttt{sin}(2k_2y)\texttt{exp}[-2\nu(k^2_1+k^2_2)t]/2k_2)$ with $u^2_0$ as the amplitude
of the force, $k_1$ and $k_2$ the corresponding wave number in $x$ and $y$ directions, and there has an analytical solution to $\bm u=(u_x, u_y)$ \cite{Guo1}
\begin{table}
\caption{Comparisons of numerical predictions by different $\delta\bm u$ at $t=2t_c$ and $x^*=0$}
{\begin{tabular}{@{}cccccc@{}} \toprule
  $\delta\bm u$ &~~~~$y^*$=-1/2 &~~~~$y^*$=-1/4  &~~~~$y^*$=0 &~~~~$y^*$=1/4 &~~~~$y^*$=1/2 \\
  \hline
  Guo \emph{et al.}\cite{Guo1} & 0.002499861& 0.001767593 & -1.457165e-016 & -0.001767593 & -0.002499861 \\
  Wagner \cite{Wagner} & 0.002499861 & 0.001767593 & -2.706163e-016 & -0.001767593 & -0.002499861\\
  $\delta\bm u1$ & 0.002499861 & 0.001767593 & 2.775552e-017 & -0.001767593 & -0.002499861\\
  $\delta\bm u2$ & 0.002499861 & 0.001767593 & -1.179610e-016 & -0.001767593 & -0.002499861 \\
  $\delta\bm u3$ & 0.002499861 & 0.001767593 & 1.248998e-016 & -0.001767593 & -0.002499861 \\
  $\delta\bm u4$ & 0.002499861 & 0.001767593 & -1.526554e-016 & -0.001767593 & -0.002499861\\
  $\delta\bm u5$ & 0.002499861 & 0.001767593 & -6.244992e-017 & -0.001767593 & -0.002499861 \\
  $\delta\bm u6$ & 0.002499861 & 0.001767593 & -2.497996e-016 & -0.001767593 & -0.002499861\\
  Eq. (\ref{Eq17}) & 0.0025 & 0.001767767 & 0 & -0.001767767 & -0.0025 \\
  \hline
\end{tabular} }
\end{table}
 \begin{eqnarray}
 \label{Eq17}
u_x(x, y, t)&=&-u_0\texttt{cos}(k_1x)\texttt{sin}(k_2y)\texttt{exp}[-\nu(k^2_1+k^2_2)t],\\
u_y(x, y, t)&=&u_0\frac{k_1}{k_2}\texttt{sin}(k_1x)\texttt{cos}(k_2y)\texttt{exp}[-\nu(k^2_1+k^2_2)t].
\label{Eq18}
\end{eqnarray}

In the simulation, the computational domain is $-\pi/2\leq x, y\leq\pi/2$, and the parameters of $u_0$, shear viscosity and wave numbers in Eqs. (\ref{Eq17}) and (\ref{Eq18}) are set to be $u_0=0.01$, $\nu=0.001$ and $k_1=k_2=1$, respectively. The results show that both numerical predictions agree well with the analytical solution at the location ($x^*=0$, $y^*$) and $t=t_c=ln2/\nu(k^2_1+k^2_2)$ and $t=2t_c$, here $x^*=x/\pi$ and $y^*=y/\pi$. In addition, the predictions with different values of $\delta\bm u$ such as $\delta\bm u=-10\delta t\bm F/\rho, -5\delta t\bm F/\rho, -\delta t\bm F/\rho,\delta t\bm F/\rho, 5\delta t\bm F/\rho$ and $10\delta t\bm F/\rho$ are compared at $t=2t_c$ and the corresponding of $\delta\bm u$ is denoted by $\delta\bm u1$ to $\delta\bm u6$ in Fig. 1, and it is observed that all the numerical predictions agree well with analytical solution.  In table 1, it shows the numerical predictions of $u_x$ as an example by different $\delta\bm u$ at $t=2t_c$ and the location $x^*=0$, $y^*=-1/2, -1/4, 0, 1/4, 1/2$ together with the analytical results for quantitative comparison. From the table, the predictions by all schemes are completely the same, which are consistent with our theoretical analysis.

\subsection{Stationary droplet}

Now a stationary droplet immersed to another fluid is simulated by pseudopotential LBE \cite{Shan,Zheng} with Carnahan-Starling (CS) equation of state (EOS) as an example to validate our force analysis, that is,
 \begin{equation}
 p_{EOS}=\rho RT\frac{1+\eta+\eta^2-\eta^3}{(1-\eta)^3}-a\rho^2,
\label{aEq8}
\end{equation}
where $T$ is the temperature, $\eta=b\rho/4$, $a$, $b$ are the constant parameters, and the internal force is given as \cite{Zheng,Zheng1}
 \begin{equation}
\bm F=-\frac{G\psi(\bm x)}{\delta t}\sum_i\omega_i\psi(\bm x+\bm{\xi}_i\delta t)\bm{\xi}_i,
\end{equation}
with $G$ and $\psi$ being the interaction strength and interaction potential respectively.
Initially, a circular droplet with a radius of R is placed in the centre of computational domain, when it reaches equilibrium state, the pressure difference $\delta p$ between inside and outside drop should satisfy the Laplace law, that is, $\delta p=\sigma/\texttt{R}$. In the simulation, a $100\times100$ mesh is used, and periodic boundary condition is applied to both directions. The model parameters are fixed as $a=1$, $R=1$, and the relaxation time $\tau=1.0$.

\begin{figure}
\center
\includegraphics[width=1\textwidth,height=0.5\textwidth]{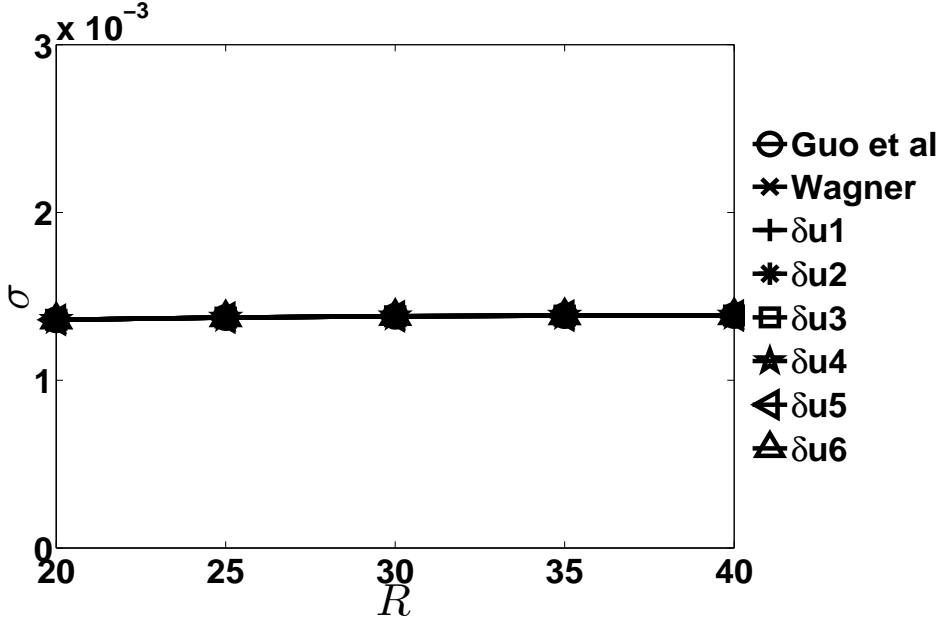}
\caption{The surface tension vs. the radius R predicted by Guo \emph{et al.},
Wagner and $\delta\bm u_1$- $\delta\bm u_6$ force schemes}\label{Fig2}
\end{figure}

In Fig. 2
, the surface tension $\sigma$ v.s. R is plotted at reduced temperature $T_r=T/T_c=0.9$, here the critical temperature $T_c=a/(10.601R)$, and it is shown that $\sigma$ is almost constant against R, this implies the Laplace law is almost satisfied. The quantitative comparisons of saturation density, $\sigma$, the magnitude of spurious velocity $V_{max}=\sqrt{u^2_x+u^2_y}$ ($u_x$ and $u_y$ the velocity components) and droplet radius predicted by the force schemes of Guo \emph{et al}, Wagner and $\delta \bm u1-\delta\bm u6$ are showed in table 2 with initial radius R=25. The results show that all force schemes could give the same predictions, and the reason is that the force schemes in table 2 satisfy the required conditions in Eq. (\ref{Eq14}).

\begin{table}[ht]
\caption{The predicted saturation density, $\sigma$, $V_{max}$ and droplet radius for stationary droplet by different force scheme}
{\begin{tabular}{@{}cccccccc@{}}
\hline
\hline
Force scheme  &$\rho_l$ &$\rho_g$& &$\sigma$ &$V_{max}$& &\texttt{R}\\
\hline
  Guo \emph{et al.} \cite{Guo1}& 0.246708196& 0.040526511 & & 0.001373667 &5.993026284e-004& & 26.2006041\\
  Wagner \cite{Wagner} & 0.246708196& 0.040526511 & & 0.001373667 &5.993026284e-004& & 26.2006041\\
  $\delta\bm u1$ & 0.246708196& 0.040526511 & & 0.001373667 &5.993026284e-004& & 26.2006041\\
  $\delta\bm u2$ & 0.246708196& 0.040526511 & & 0.001373667 &5.993026284e-004& & 26.2006041\\
  $\delta\bm u3$ & 0.246708196& 0.040526511 & & 0.001373667 &5.993026284e-004& & 26.2006041\\
  $\delta\bm u4$ & 0.246708196& 0.040526511 & & 0.001373667 &5.993026284e-004& & 26.2006041\\
  $\delta\bm u5$ & 0.246708196& 0.040526511 & & 0.001373667 &5.993026284e-004& & 26.2006041\\
  $\delta\bm u6$ & 0.246708196& 0.040526511 & & 0.001373667 &5.993026284e-004& & 26.2006041\\
  \hline
\end{tabular} }
\end{table}

\subsection{Droplet on wettability surface }
\begin{figure}
\center
\includegraphics[width=0.45\textwidth,height=0.225\textwidth]{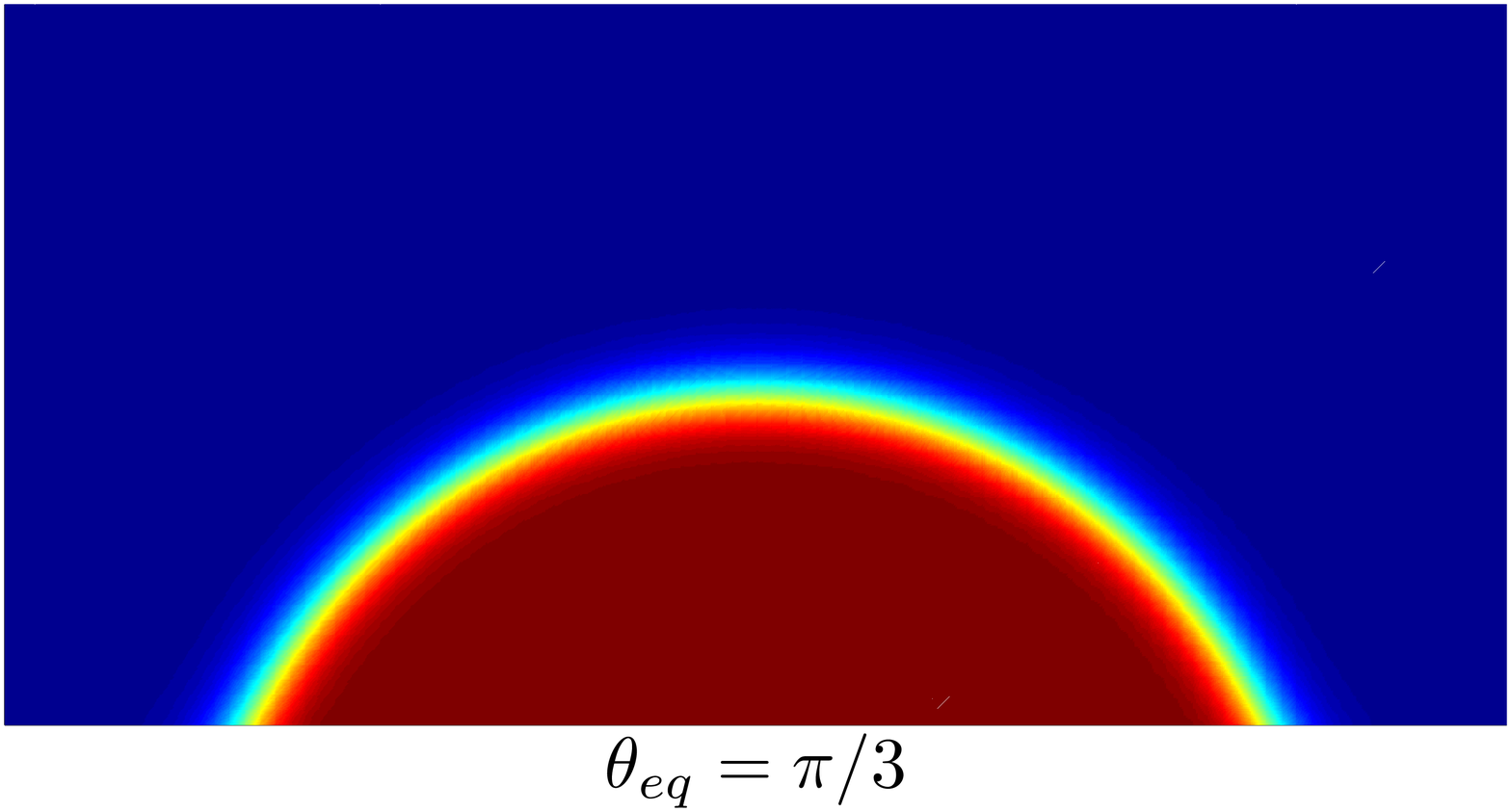}
\includegraphics[width=0.45\textwidth,height=0.225\textwidth]{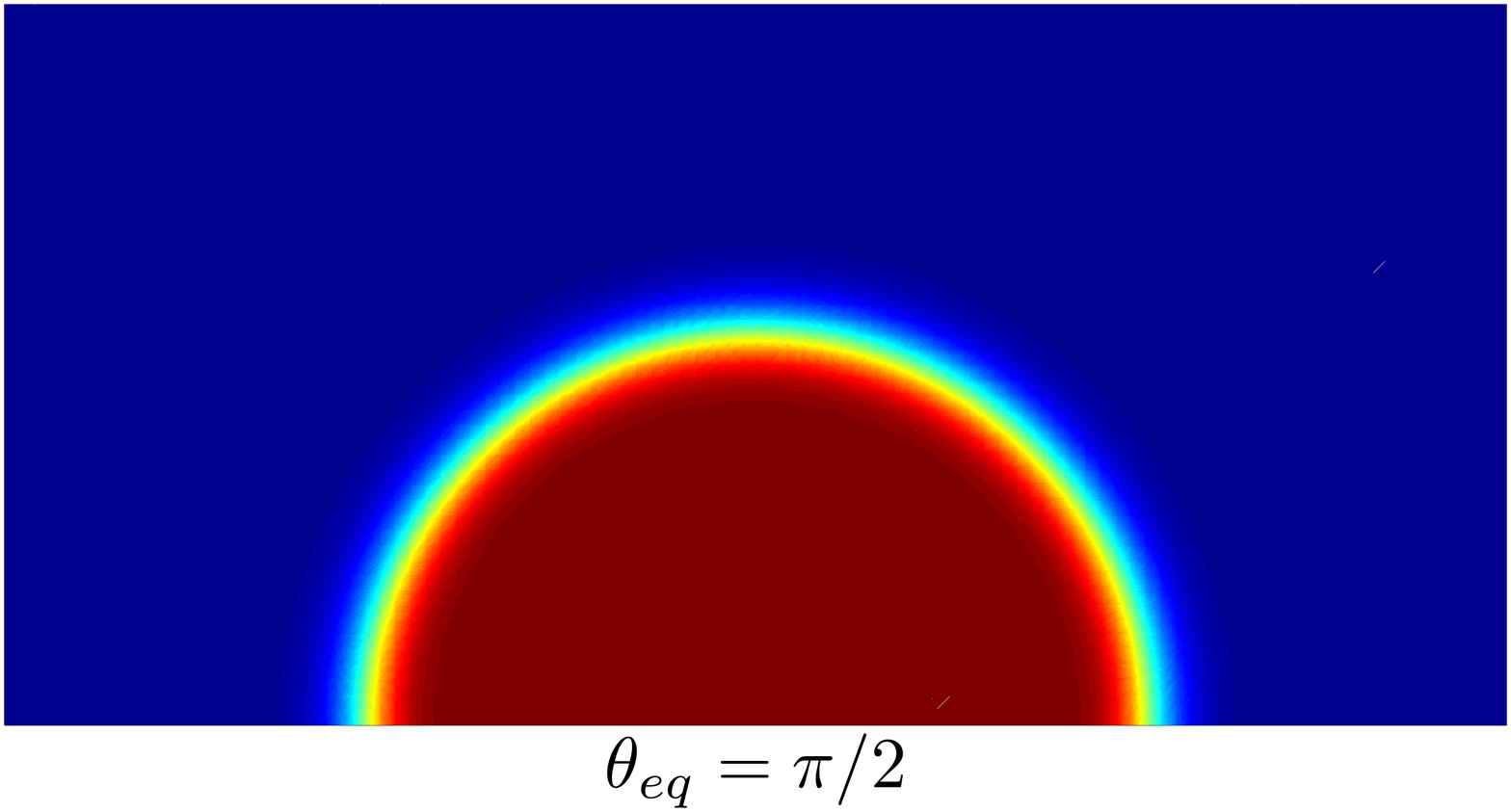}
\includegraphics[width=0.45\textwidth,height=0.225\textwidth]{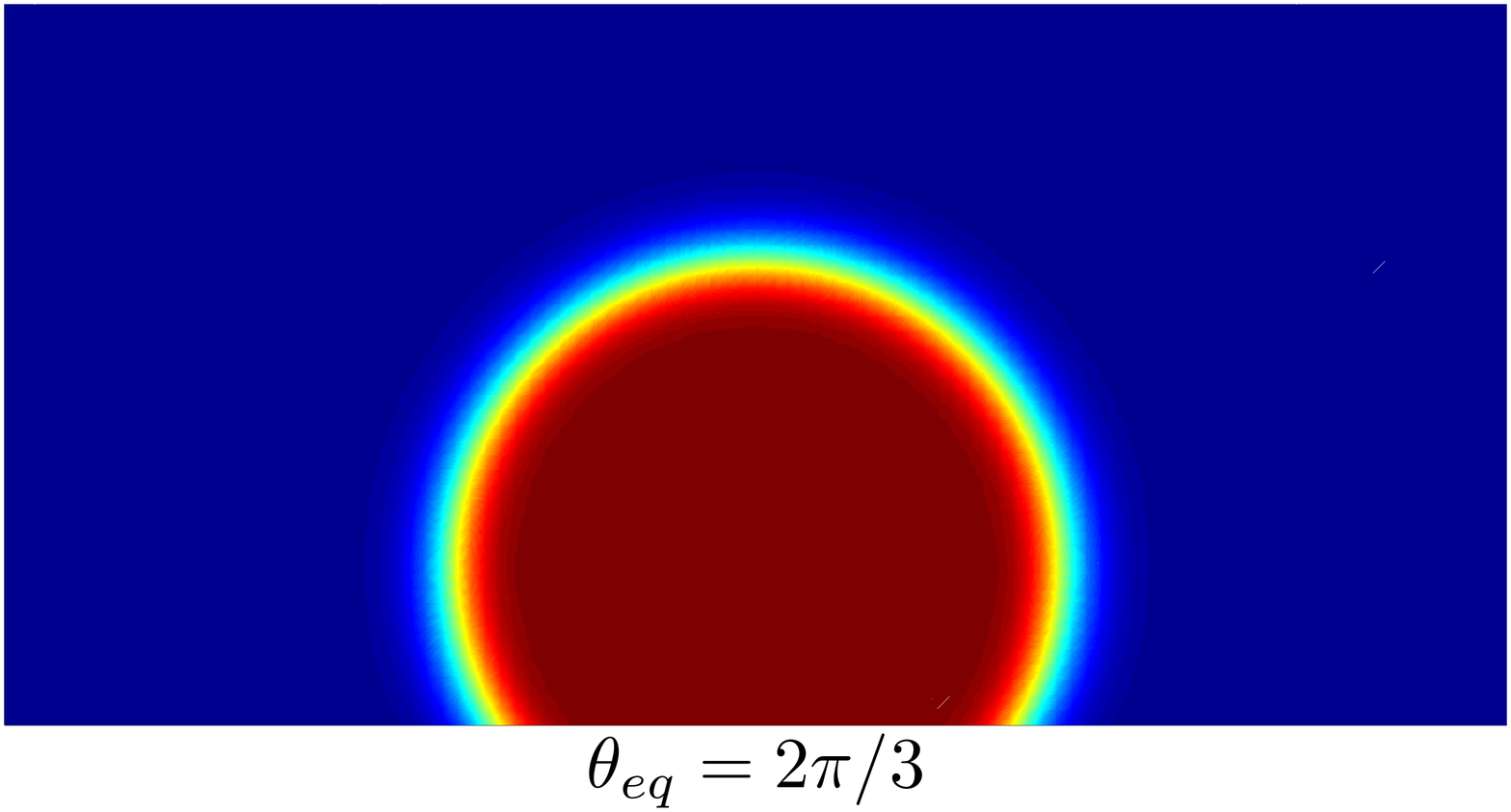}
\caption{The density contours of droplet on solid surface with $\theta_{eq}=\pi/3, \pi/2$ and $2\pi/3$ by LBE}\label{Fig3}
\end{figure}
Finally, a droplet on a wettability surface is also investigated by the pseudopotential LBE with CS EOS at $T_r=0.9$. In the simulation, a $100\times 50$ mesh is applied, periodic boundary condition is applied to $x$ direction and nonslip boundary on solid surface. Similar to the geometric formulation in Ref. \cite{Ding} is used to mimic the interaction between the fluid molecules and the solid surface, which is given as
\begin{equation}
 \psi_{\varsigma,0}=\psi_{\varsigma,2}+\texttt{tan}(\frac{\pi}{2}-\theta)|\psi_{\varsigma+1,1}-\psi_{\varsigma-1,1}|
\end{equation}
where $\varsigma$ is the coordinates along the solid surface, 0, 1 and 2 are respectively the ghost nodes,
boundary nodes and fluid nodes normal to the solid boundary, and $\theta$ is the contact angle.

Initially, a half droplet is placed on the solid surface with a constant volume $V=\pi \texttt{R}^2/2$. The density contours of droplet on the solid surface with equilibrium contact angle $\theta_{eq}= \pi/3, \pi/2$ and $2\pi/3$ are showed in Fig. 3
, and the results show that present fluid-solid interaction treatment could capture the effect of wettability. The numerical predictions of saturation density, contact angle $\theta$, the magnitude of spurious velocity $V_{max}$ by aforementioned force schemes are compared with $\theta_{eq}=2\pi/3$ in table 3, it is observed that all the numerical predictions are also consistent with each other. This implies that the constraints of the force term in Eq. (\ref{Eq14}) should be satisfied with the apparently relaxation process in LBE to derive correct hydrodynamic equations at NS level.
\begin{table}[ht]
\caption{The predicted saturation density, $\theta$, and $V_{max}$ by different force scheme}
{\begin{tabular}{@{}cccccccc@{}}
\hline
\hline
Force scheme  &$\rho_l$ &$\rho_g$& &$\theta$ &$V_{max}$& \\
\hline
  Guo \emph{et al.} \cite{Guo1}& 0.247052601& 0.040714200 & & 120.483706 &0.014162536& \\
  Wagner \cite{Wagner}&  0.247052601& 0.040714200 & & 120.483706 &0.014162536& \\
  $\delta\bm u1$ &  0.247052601& 0.040714200 & & 120.483706 &0.014162536& \\
  $\delta\bm u2$ & 0.247052601& 0.040714200 & & 120.483706 &0.014162536& \\
  $\delta\bm u3$ &  0.247052601& 0.040714200 & & 120.483706 &0.014162536& \\
  $\delta\bm u4$ &  0.247052601& 0.040714200 & & 120.483706 &0.014162536& \\
  $\delta\bm u5$ &  0.247052601& 0.040714200 & & 120.483706 &0.014162536& \\
  $\delta\bm u6$ &  0.247052601& 0.040714200 & & 120.483706 &0.014162536& \\
  \hline
\end{tabular} }
\end{table}

\section{Conclusion}

Theoretical analysis suggests that LBE with a force term is the way to approximate
the kinetic Boltzmann equation, and the CE analysis shows that both $f^{(0)}$ and local equilibrium density distribution function are $f^{(eq)}(\rho, \bm u,T)$ with/without a driven force. Our results also reveal that LBE with a given state relaxation process should be consistent with the kinetic theory, this requires $f^{(0)}_i$ and local equilibrium density distribution function must to be $f^{(eq)}_i(\rho, \bm u)$, and the general force $F_i$ should satisfy Eq. (\ref{Eq14}), by which all the predictions are consistent with each other. It should be pointed out that Wagner \cite{Wagner} analyzed the difference between the force schemes of Guo \emph{et al.} and Wagner, the discussion showed that the correct term in the force term was different even without the higher order interfacial term in Wagner's scheme ( See Eqs. (59) and (62) in Ref.\cite{Wagner} ). He argued that this inconsistency was arisen from the different expansion technique, \emph{i.e.}, the CE analysis was applied in Guo \emph{et al.} force scheme while Taylor expansion technique was used in Wagner's force scheme. However, present work shows both force schemes are completely the same without the higher order interfacial term in Wagner's force scheme, and the reason for this inconsistency is that the given state $f^{(eq)}_i(\rho, \bm u^*)$  is not expanded around physical local equilibrium state $f^{(eq)}_i(\rho, \bm u)$ by CE analysis in Ref. \cite{Guo1}

Subsequently, we compare the predictions by different $\delta\bm u$ with the analytical solution, and the internal force interaction in two phase flow such as  stationary droplet problem and the droplet on wettability surface are investigated, all predictions are completely the same or in good agreement with analytical one, which are consistent with our theoretical analysis. This provides a way to model and develop numerical methods for external/internal force driven flows.

\section*{Acknowledgments}
This work is supported from the Natural Science Foundation of China (Grant No. 51506097) and Anhui Provincial Natural Science research project of China (Grant No. KJ2015A209)


\begin{thebibliography}{99}
\bibitem{Cercignani}C. Cercignani, Theory and Application of the Boltzmann Equation, Scottish Academic Press, Edinburgh, 1975.
\bibitem{BGK}P. L. Bhatnagar, E. P. Gross, and M. Krook, A Model for Collision Processes in Gases. I. Small Amplitude Processes in Charged and Neutral One-Component Systems, Phys. Rev. 94(3) (1954) 511-525.
\bibitem{Succi}S. Succi, The Lattice Boltzmann Equation for Fluid Dynamics and Beyond, Oxford University Press, Oxford, 2001.
\bibitem{GuoB}Z. L. Guo and C. Shu, Lattice Boltzmann Method and Its Applications in Engineering, World Scientific, Singapore, 2013.
\bibitem{Xu}K. Xu, L. Martinelli, and A. Jameson, Gas-Kinetic Finite Volume Methods, Flux-Vector Splitting, and Artificial Diffusion, J. Comput. Phys. 120(1) (1995) 48-65.
\bibitem{Aidun}C. K. Aidun and J. R. Clausen, Lattice-Boltzmann Method for Complex Flows, Annu. Rev. Fluid Mech. 42(1) (2010) 439-472.
\bibitem{Mehr}M. Mehravaran and S. K. Hannani, Simulation of incompressible two-phase flows with large density differences employing lattice Boltzmann and level set methods, Comput. Methods Appl. Mech. Engrg. 198(2) (2008) 223-233.
\bibitem{Torres}S. A. Galindo-Torres, A coupled Discrete Element Lattice Boltzmann Method for the simulation of fluid-solid interaction with particles of general shapes, Comput. Methods Appl. Mech. Engrg. 265(2) (2013) 107-119.
\bibitem{Shan}X. Shan and H. Chen, Lattice Boltzmann model for simulating flows with multiple phases and components, Phys. Rev. E 47(3) (1993) 1815-1819.
\bibitem{Shan1}X. Shan and H. Chen, Simulation of nonideal gases and liquid-gas phase transitions by the lattice Boltzmann equation, Phys. Rev. E 49(4) (1994) 2941-2948.
\bibitem{He}X. He, S. Chen,  G. D. Doolen, A novel thermal model for the lattice Boltzmann method in incompressible limit, J. Comput. Phys. 146(1) (1998) 282-300.
\bibitem{Ladd} A. J. C. Ladd and R. Verberg, Lattice-Boltzmann simulations of particle-fluid suspensions, J. Stat. Phys. 104(5) (2001) 1191-1251.
\bibitem{Guo1} Z. L. Guo, C. G. Zheng, and B. C. Shi, Discrete lattice effects on the forcing term in the lattice Boltzmann method, Phys. Rev. E 65(4) (2002) 046308.
\bibitem{Wagner}A. J. Wagner, Thermodynamic consistency of liquid-gas lattice Boltzmann simulations, Phys. Rev. E 74(2) (2006) 056703.
\bibitem{Kuper}A. L. Kupershtokh, D. A. Medvedev, and D. I. Karpov, On equations of state in a lattice Boltzmann method, Comput. Math. Appl. 58(5) (2009) 965-974.
\bibitem{Zheng}L. Zheng, Q. L. Zhai and S. Zheng, Analysis of force treatment in the pseudopotential lattice Boltzmann equation method, Phys. Rev. E 95(4) (2017) 043301.
\bibitem{Zheng1}Q. L. Zhai, L. Zheng and S. Zheng, Pseudopotential lattice Boltzmann equation method for two-phase flow: A higher-order Chapmann-Enskog expansion, Phys. Rev. E 95(2-1) (2017) 023313.
\bibitem{Ding}H. Ding and P. D. M. Spelt, Wetting condition in diffuse interface simulations of contact line motion, Phys. Rev. E 75(2) (2007) 046708.
\end{thebibliography}
\end{document}